\documentclass[aps,prb,twocolumn,superscriptaddress,showpacs,floatfix]{revtex4}
\usepackage{graphicx}
\usepackage{dcolumn}
\usepackage{bm}
\usepackage{stmaryrd}
\usepackage{latexsym}
\usepackage{amssymb}
\usepackage{amsfonts}
\usepackage{amsmath}

\begin{document}
\title{Fulde-Ferrell-Larkin-Ovchinnikov superfluidity in one-dimensional optical lattices}

\author{M. Rizzi}
\affiliation{Max-Planck-Institut f{\"u}r Quantenoptik,
             Hans-Kopfermann-Str. 1, D-85748 Garching, Germany}
\affiliation{NEST-CNR-INFM and Scuola Normale Superiore, I-56126 Pisa, Italy}
\author{Marco Polini}
\email{m.polini@sns.it}
\affiliation{NEST-CNR-INFM and Scuola Normale Superiore, I-56126 Pisa, Italy}
\author{M.A. Cazalilla}
\affiliation{Centro de F\'isica de Materiales, Centro Mixto CSIC-UPV/EHU, P.
             Manuel de Lardizabal 3, 20018, San Sebasti\'an, Spain}
\affiliation{Donostia International Physics Center (DIPC),
             Manuel de Lardizabal 4, 20018, San Sebastian, Spain}
\author{M.R. Bakhtiari}
\affiliation{Nanoscience Center, Department of Physics, University of Jyv\"askyl\"a, 
             P.O. Box 35, 40014, Jyv\"askyl\"a, Finland}
\author{M.P. Tosi}
\affiliation{NEST-CNR-INFM and Scuola Normale Superiore, I-56126 Pisa, Italy}
\author{Rosario Fazio}
\affiliation{International School for Advanced Studies (SISSA), 
             via Beirut 2-4, I-34014 Trieste, Italy}
\affiliation{NEST-CNR-INFM and Scuola Normale Superiore, I-56126 Pisa, Italy}

\date{\today}
\begin{abstract}
Spin-polarized attractive Fermi gases in one-dimensional (1D) optical lattices 
are expected to be remarkably good candidates for the observation of the Fulde-Ferrell-Larkin-Ovchinnikov 
(FFLO) phase.  We model these systems with an attractive Hubbard model with population 
imbalance.  By means of the density-matrix renormalization-group method we compute
the pairing correlations as well as the static spin and charge structure factors 
in the whole range from weak to strong coupling.  We demonstrate that 
pairing correlations exhibit quasi-long range order and oscillations at the wave number expected from
FFLO theory. However, we also show by numerically computing the mixed 
spin-charge static structure factor that charge and spin degrees of freedom appear to be coupled already 
for small imbalance. We discuss the consequences of  this coupling for the observation of the FFLO phase,
as well as for the stabilization of the quasi-long range order into long-range order  by coupling 
many identical 1D systems, as in quasi-1D optical lattices.
\end{abstract} 

\maketitle

\section{Introduction}

Multi-component attractive fermionic systems with unequal masses, densities or chemical 
potentials have attracted continued interest for many decades in several 
fields of physics ranging from high-energy~\cite{rajagopal_2002,casalbuoni_2004} to
condensed matter~\cite{casalbuoni_2004,yang_review_2006} and, more recently, atomic 
physics~\cite{population_imbalance,phase_contrast,parish_2007}. 
The interplay between pairing and density imbalance of the different fermion 
species leads to a rich scenario, which includes the possibility of 
various exotic superconducting states~\cite{exotic}. 
In this context, the Fulde-Ferrell-Larkin-Ovchinnikov (FFLO) phase~\cite{FFLO} has recently 
attracted a great deal of interest from both the experimental and the 
theoretical community~\cite{rajagopal_2002,casalbuoni_2004,yang_review_2006,
population_imbalance,phase_contrast,parish_2007}.  
%In two or three
%dimensional (3D) systems exhibiting this phase,  
In the FFLO phase Cooper pairing occurs between a 
fermion with momentum ${\bm k}$ and spin $\uparrow$ and a fermion with momentum 
$-{\bm k}+{\bm q}$ (${\bm q} \neq {\bm 0}$) and spin $\downarrow$.
As a result, the superconducting order parameter becomes spatially dependent. 
Originally, the most favorable  systems for the observation of the FFLO  phase were predicted to be 
clean superconducting films in the presence of an in-plane (\emph{i.e.} Zeeman) magnetic field,  
above the so-called Clogston-Chandrasekhar limit~\cite{chandrasekhar_clogston}. Nevertheless, despite the fact that
the original  prediction dates back to more than thirty years ago, 
the FFLO phase has been very elusive to detect.  

The experimental realization of interacting trapped Fermi gases
with population imbalance~\cite{population_imbalance,phase_contrast} has  
renewed the hope of observing the FFLO, thus stimulating an 
intense theoretical activity~\cite{parish_2007,population_imbalance_theory}. So far, most of 
the theoretical analysis has focused on 3D cold atomic systems. 
However, as in the case of solid-state superconductors, 
the region of phase diagram where the FFLO phase has been found to be stable is 
quite small~\cite{casalbuoni_2004,parish_2007}. On the other hand, quasi-one dimensional 
or strongly anisotropic systems (such as coupled chains, heavy-fermion, organic, 
high-$T_{\rm c}$, and ${\rm CeCoIn}_5$ superconductors) are believed 
to be good candidates for the realization of the FFLO phase~\cite{casalbuoni_2004,
yang_review_2006,pruschke,yang}. Since the dimensionality of cold atomic systems
can be easily tuned, and indeed cold atoms have already been successfully trapped in 1D 
geometries~\cite{cold_atoms_low_d}, it seems natural to consider these low dimensional systems as
the ideal candidates to observe non-homogeneous superconductivity of the FFLO type. 

Many important results are available on the properties of
spin-polarized 1D Fermi  systems with attractive interactions,
which have been obtained by different methods and techniques. These include
the Bethe-{\it Ansatz} solutions of certain exactly solvable models like the 
Hubbard or Gaudin-Yang models~\cite{Bahder86,Guan07,hui_hu_prl_2007,orso_prl_2007},
as well as different types of numerical approaches such as density-matrix renormalization-group (DMRG)~\cite{Roux07,Feiguin07,Tezuka07} and quantum Monte Carlo (QMC)~\cite{Batrouni07}, or  field theoretical 
techniques like bosonization~\cite{yang,Roux07}.  
Very recently
%, in the context of cold atomic gases, 
Orso~\cite{orso_prl_2007} and Hu {\it et al.}~\cite{hui_hu_prl_2007} have studied
the phase diagram of harmonically-trapped 1D polarized Fermi gases  by combining  the exact solution
of the Gaudin-Yang model  with a local-density approximation. Mean-field theory has also been applied by
Liu {\it et al.}~\cite{hui_hu_pra07}, although it is known that it has a number of 
limitations~\cite{Marsiglio97} in 1D, particularly as far as paring correlations are concerned.
DMRG has been employed by Feiguin {\it et al.}~\cite{Feiguin07} and  
Tezuka and Ueda~\cite{Tezuka07}, and QMC by Batrouni {\it et al.}~\cite{Batrouni07} 
to investigate the pairing correlations in the spin-polarized ground state 
of the attractive Hubbard model in the presence of harmonic trapping. 
Previously, Yang~\cite{yang} used  bosonization to study the pairing correlations and the phase diagram of a single 
1D Fermi system as well as an array of weakly-coupled 1D Fermi systems in the presence of a Zeeman field.

Yang's analysis is valid only close to a continuous magnetic-field-driven transition 
from a uniform BCS phase and an FFLO phase, 
which he assumed to belong to the commensurate-incommensurate  universality class~\cite{cic}.  
Another important assumption in Ref.~\onlinecite{yang} is that 
charge and spin degrees of freedom are decoupled  at low energies for small polarization. 
However, this scenario does not apply to the Hubbard model away from half-filling~\cite{Bahder86,Woynarovich,Frahm07} 
nor to the Gaudin-Yang model~\cite{Guan07}, which are the relevant models for current 1D cold atomic systems. 
As we show in this work, charge and spin degrees of freedom are indeed coupled 
already for small polarization, which leads to important differences as compared to the
scenario described by Yang.  We also numerically demonstrate
that the pairing correlation function  exhibits  prominent 
oscillations with a wave number equal  (up to finite size corrections, see below) to the
difference of Fermi wave numbers, $q_{\rm FFLO}=|k_{{\rm F}\uparrow} - k_{{\rm F}\downarrow}|$, as predicted by FFLO
theory in 1D~\cite{yang} and in agreement with a number of  Luttinger-type theorems~\cite{Haldane92,YOA}.   
Thus, the finite-wave-number oscillations in the pairing correlation
function  can be regarded as due to the excess of $n_{\uparrow}-n_{\downarrow}$ unpaired majority-spin 
fermions~\cite{yang}. This is because in 1D the Fermi wave number and the density are 
proportional  to each other: $k_{{\rm F}\sigma} = \pi n_{\sigma}$.  On the other hand,
this relationship is no longer linear in dimensionality higher than one, and 
in this case the oscillations in the order parameter are related to the center-of-mass  
momentum ${\bm q}$ of the Coopers pairs.
%Therefore, in a certain sense, 
These observations seem to indicate that in the 1D case it is not entirely clear whether there is a 
strict close parallelism 
with higher dimensional FFLO, and in some respects the system can be also understood as a a coupled 
Bose-Fermi mixture of spin-singlet pairs (the bosons) and unpaired fermions~\cite{Woynarovich,hui_hu_prl_2007}. 

The paper is organized as follows. Section~\ref{sect:model} presents the  Hamiltonian that 
we use to describe the system of physical interest, while Section~\ref{sect:results} reports and 
discusses our main numerical results. Our main conclusions are briefly reported 
in Section~\ref{sect:summary}.

\section{The model}
\label{sect:model}

 We consider a two-component mixture with a total of 
$N$ fermionic atoms loaded in a 1D optical lattice of $L$ 
 sites (the lattice constant is taken to be unity). The fermions are assumed
to interact {\it via} attractive on-site interactions, whose strength can be tuned 
\emph{e.g.} by means of a Feshbach resonance.  Sufficiently away from resonance(s), 
this system is modeled by the  attractive Hubbard model:
\begin{equation}\label{eq:hubbard}
	{\hat {\cal H}}=-t\sum_{\sigma,\ell=1}^{L-1}({\hat c}^{\dagger}_{\ell\sigma}
	{\hat c}_{\ell+1\sigma}+{\rm H}.{\rm c}.)-U\sum_{\ell=1}^{L}\,
        {\hat n}_{\ell\uparrow}{\hat n}_{\ell\downarrow}\,,
\end{equation}
where $t$ is the hopping parameter, ${\hat c}^{\dagger}_{\ell\sigma}$ (${\hat c}_{\ell\sigma}$) is 
the creation (destruction)
fermion operator in the $\ell$-site ($\ell\in[1,L]$),
$\sigma=\uparrow,\downarrow$ the pseudospin-$1/2$ index 
(in experiments this labels the two different atomic hyperfine states of the mixture), 
$U>0$ is the strength of the on-site Hubbard attraction,
${\hat n}_{\ell\sigma}={\hat c}^{\dagger}_{\ell\sigma}{\hat c}_{\ell\sigma}$, and $N_{\sigma}=
\sum_{\sigma} {\hat n}_{\ell\sigma}$.  In order to simulate the effect of an external trapping potential, 
open-boundary conditions (OBC) breaking translational symmetry will be used (these are indeed the most suitable
conditions for the DMRG treatment of the above model~\cite{dmrg}). Our calculations are performed 
in the canonical 
ensemble, and the results apply only to lattices away from half-filling, that is, when
$N \neq L$.  In the calculations, the spin polarization $\delta=(N_\uparrow-N_\downarrow)/(N_\uparrow+N_\downarrow)$ 
was varied by decreasing $N_{\downarrow}$ 
while keeping  constant the number of ``background''
up-spin atoms $N_\uparrow$, from $N_\downarrow=N_\uparrow$ (the unpolarized case, {\it i.e.} $\delta=0$) 
all the way down to $N_\downarrow=0$ 
(the fully polarized case, {\it i.e.} $\delta=1$). 

In the unpolarized case ($\delta = 0$), all fermions pair into spin singlets due to 
the attractive on-site interaction.  This yields a gap to all spin excitations and 
therefore spin-spin correlations decay exponentially with distance. 
Singlet superconducting and charge-density wave correlations exhibit a slower decay (of 
power-law type in the ground state of a thermodynamically
large system), being singlet superconducting
correlations the ones that dominate at long distances in systems away from half-filing~\cite{giamarchi_book}. 
The aim of this work is to study the nature of superfluidity for 
$0 <  \delta< 1$ as a function of the dimensionless ratio $U/t$ 
(in the fully polarized case, where $\delta=1$, ${\hat {\cal H}}$ describes a system 
of $N = N_\uparrow$ noninteracting fermions). The expectation values 
$\langle...\rangle$ of all operators below are understood to be taken over the ground state of 
${\hat {\cal H}}$.
%, which has been obtained by means of the DMRG algorithm~\cite{dmrg}. 
%When properly converged, the latter yields essentially exact results for an arbitrary value of $U/t$.
%

\section{Numerical results and discussion}
\label{sect:results}

\begin{figure}
\begin{center}
\tabcolsep=0cm
\begin{tabular}{c}
\includegraphics[width=1.00\linewidth]{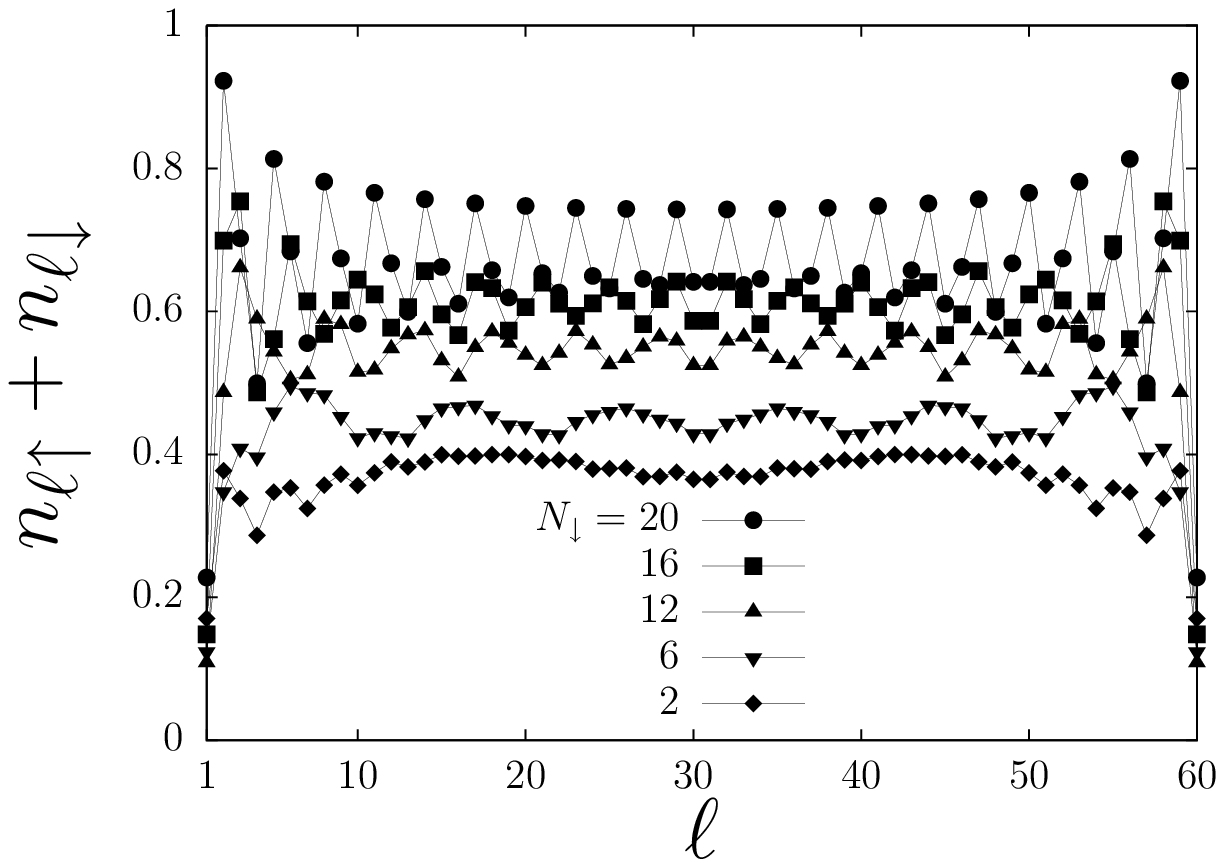}\\
\includegraphics[width=1.00\linewidth]{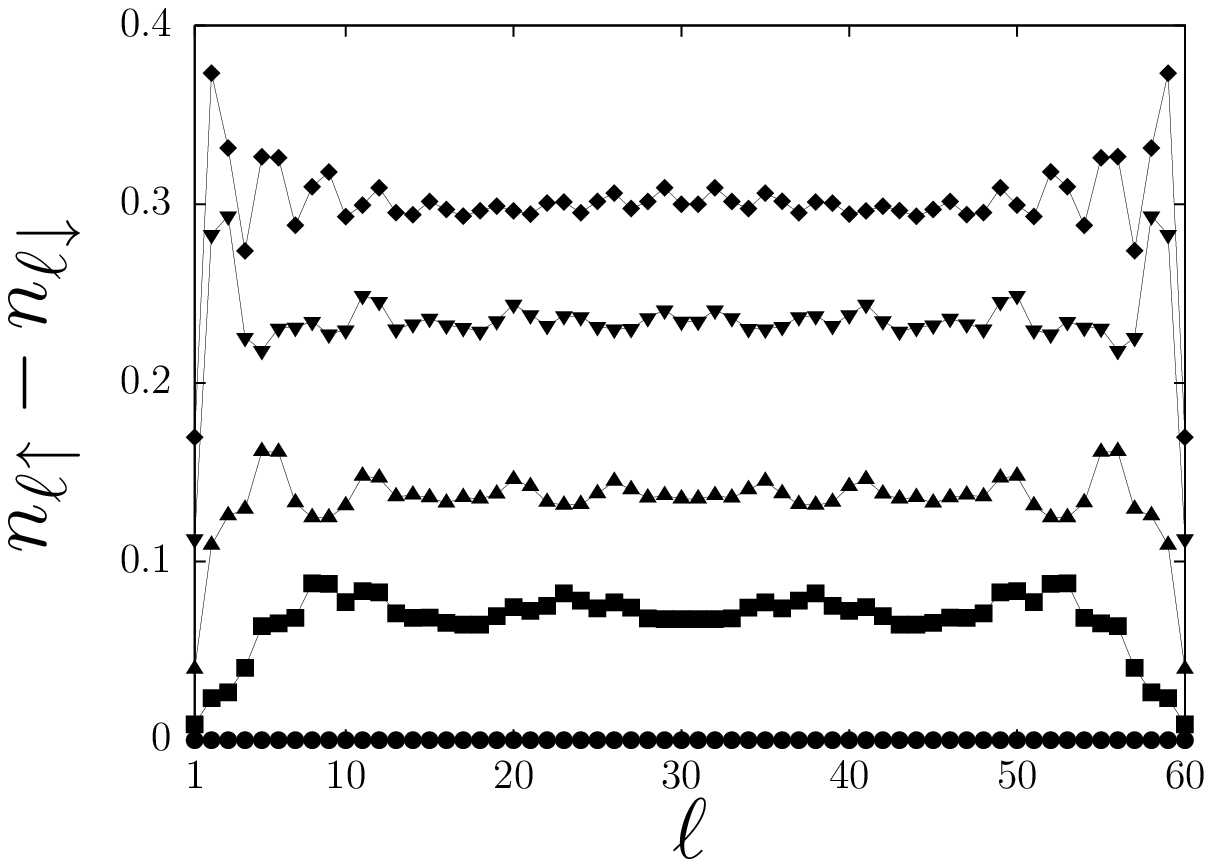}
\end{tabular}
\caption{Top panel: DMRG results for the total site occupation $n_{\ell\uparrow}+n_{\ell\downarrow}$ 
	as a function of site position $\ell$ for a system with $N_\uparrow=20$ fermions 
	in $L=60$ lattice sites and $U/t=5$. 
	The number of down-spin fermions is $N_\downarrow=20, 16, 12, 6$ and $2$ 
	(the corresponding spin polarization is $\delta=0\%, 11\%, 25\%, 54\%$, and $82\%$). 
        Bottom panel: the local spin polarization $n_{\ell\uparrow}-n_{\ell\downarrow}$ as a function of $\ell$ 
        for the same system parameters as in the top panel.
	The thin solid lines are just guides to the eye.\label{fig:one}}
\end{center}
\end{figure}

Due to the OBC (or, in general,  to any  external potential that 
breaks the Bloch translational invariance of the lattice) the spin-resolved  
site occupation profiles, $n_{\ell \sigma}=\langle {\hat n}_{\ell \sigma}\rangle$, 
exhibit Friedel oscillations. This is
illustrated in Fig.~\ref{fig:one}. In the unpolarized $\delta=0$ case the Friedel oscillations 
in $n_\uparrow$ are in-phase with those in  $n_\downarrow$ giving rise to large-amplitude atomic-density 
waves~\cite{gao_prl_2007,karim_pour_2007} in the total site occupation 
$n_{\ell\uparrow}+n_{\ell\downarrow}$. 
As it is clear from the top panel of Fig.~\ref{fig:one}, 
in the general $\delta \neq 0$ case the total site occupation displays $N_\downarrow$ 
maxima associated with the formation of $N_\downarrow$ spin-singlet pairs that are 
delocalized over the lattice. In the bottom panel of Fig.~\ref{fig:one}, we show 
the  local spin polarization $n_{\ell\uparrow}-n_{\ell\downarrow}$ 
(which could be measured through phase-sensitive optical imaging~\cite{phase_contrast}). For small 
$\delta$ (see {\it e.g.} the plot for $N_\uparrow=20$ and $N_\downarrow=16$) the local spin polarization displays 
$N_\uparrow-N_\downarrow$ maxima corresponding to the number of fermions 
that are left unpaired. With increasing $\delta$ though the spatial dependence of the local spin 
polarization becomes more complicated: the amplitude of the oscillations in the bulk becomes indeed 
smaller, thus making it hard to clearly identify $N_\uparrow-N_\downarrow$ maxima. These, however, 
are not distinctive and unambiguous signals of FFLO pairing.

\begin{figure}
\begin{center}
\tabcolsep=0cm
\begin{tabular}{c}
\includegraphics[width=1.00\linewidth]{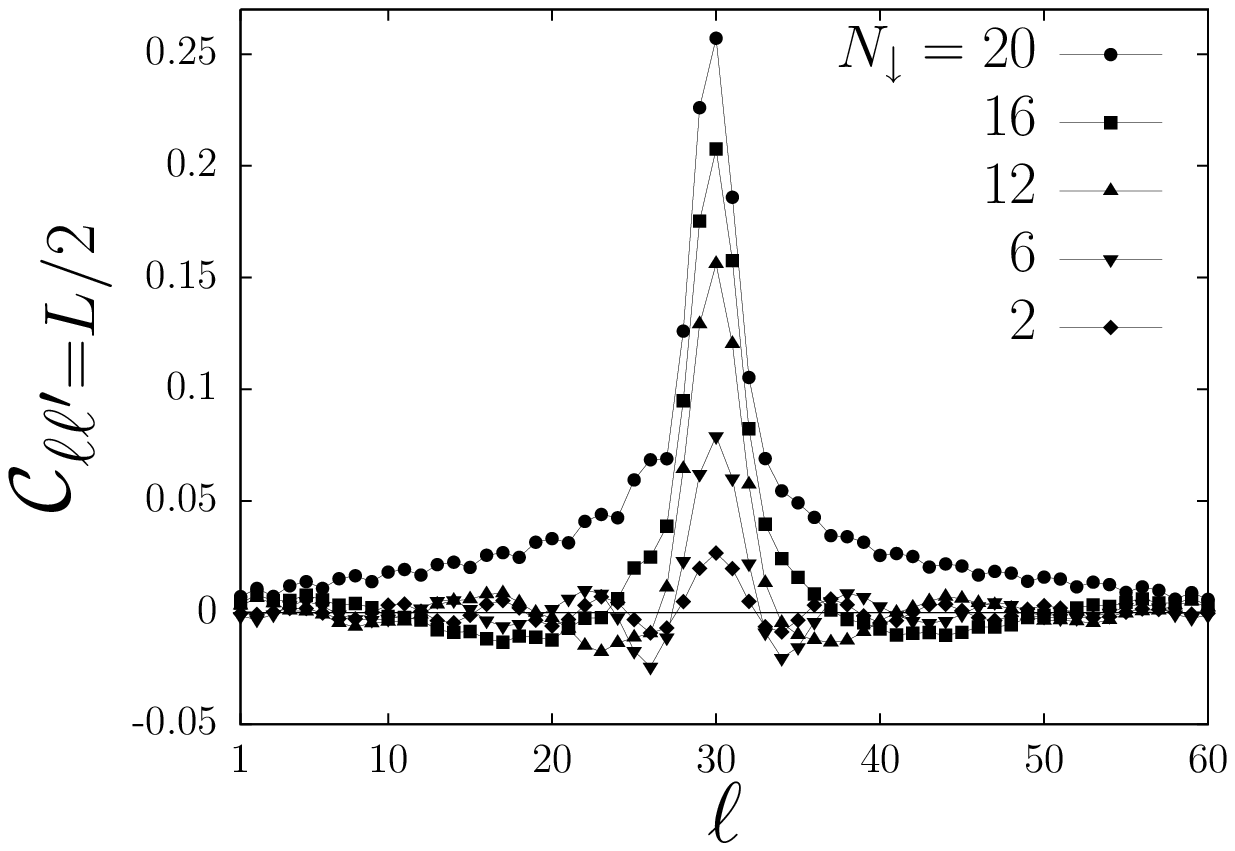}\\
\includegraphics[width=1.00\linewidth]{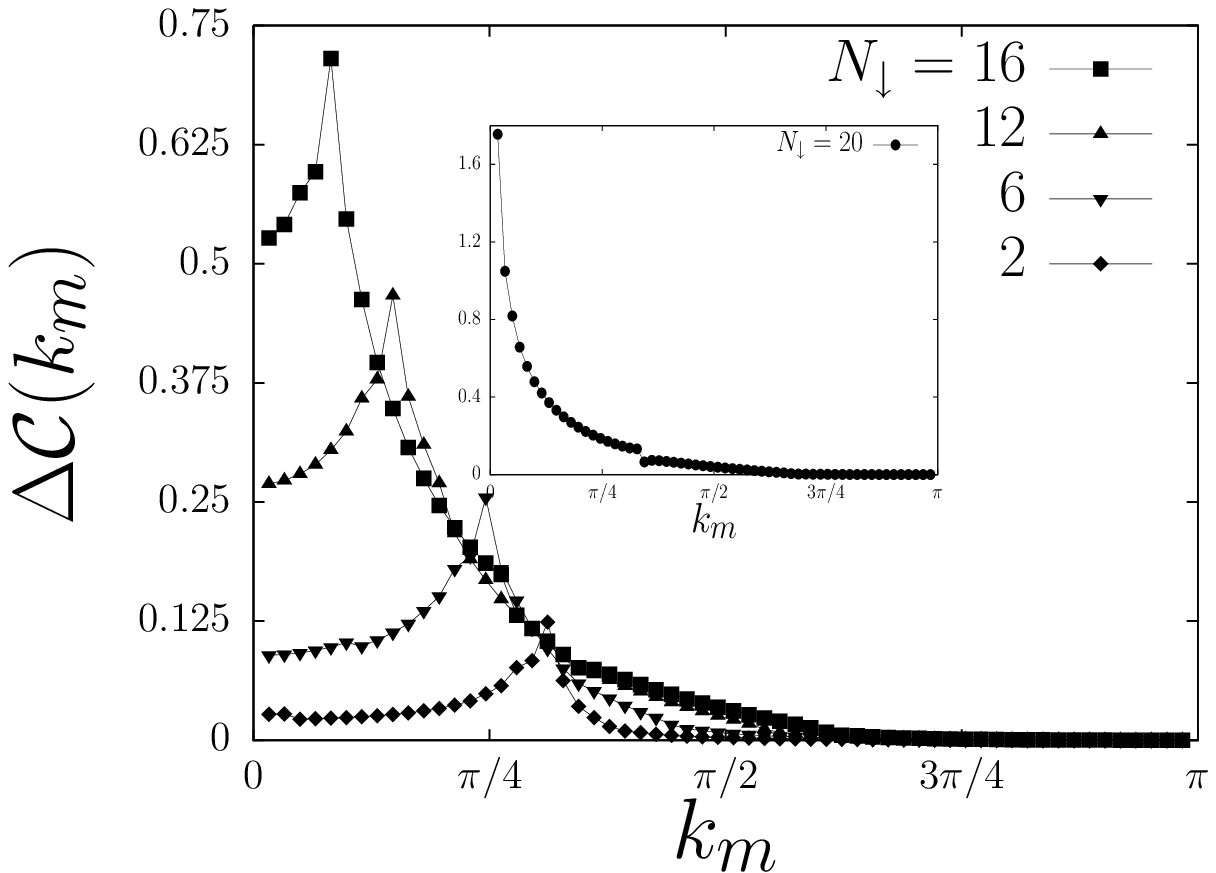}
\end{tabular}
\caption{Top panel: DMRG results for the pairing correlator 
        ${\cal C}_{\ell\ell'=L/2}$ as a function of site position $\ell$ 
	for a system with $N_\uparrow=20$ fermions in $L=60$ lattice sites and $U/t=5$. 
	The number of down-spin fermions is $N_\downarrow=20, 16, 12, 6$ and $2$ 
	(the corresponding spin polarization is $\delta=0\%, 11\%, 25\%, 54\%$, and $82\%$). 
	Bottom panel: Interaction contribution to the momentum-space pairing correlator
	$\Delta {\cal C}(k_m)$ as a function of wave number $k_m=\pi m/(L+1)$. The symbol 
	coding is the same as in the top panel. Note the well-defined peaks at $q_{\rm FFLO}$. 
	In the inset we show $\Delta {\cal C}(k_m)$ for the unpolarized system 
        with $N_\uparrow=N_\downarrow=20$, which shows a narrow peak at $k_1$.
	The thin solid lines are just guides to the eye.\label{fig:two}}
\end{center}
\end{figure}

We thus proceed below to present a study of pairing correlations: 
the model described in Eq.~(\ref{eq:hubbard}), in fact, cannot sustain any 
true long-range order~\cite{giamarchi_book} in 1D, {\it i.e.} 
the ground-state expectation value of the pairing operator 
${\hat \Delta}_\ell={\hat c}_{\ell\downarrow}{\hat c}_{\ell\uparrow}$ is zero.
In the unpolarized case and for an extended system, the correlation function of the pairing 
operator ${\cal C}_{\ell\ell'}=\langle {\hat \Delta}^\dagger_\ell{\hat \Delta}_{\ell'} \rangle$
decays with a power law $|\ell-\ell'|^{-1/K_\rho}$ at large distances, where $1\le K_\rho \le 2$ 
is an interaction-dependent Luttinger-liquid dimensionless parameter~\cite{giamarchi_book}. In the top panel 
of Fig.~\ref{fig:two} we illustrate our DMRG results for the spin-polarization dependence 
of ${\cal C}_{\ell\ell'=L/2}$ at $U/t=5$, which measures real-space superfluid correlations 
between the site $\ell'=L/2$ (the center of the trap) and all the other sites. 
For $\delta=0$ the power-law decay of the ${\cal C}_{\ell\ell'=L/2}$ for $|\ell-L/2|\gg 1$ 
is clearly visible. For finite $\delta$, 
instead, the superfluid correlator is characterized by a distinctive oscillatory character~\cite{yang} 
and a very simple nodal structure with exactly $N_\uparrow-N_\downarrow$ zeroes. We have carefully checked that 
the long-distance decay of ${\cal C}_{\ell\ell'=L/2}$ 
is still power-law, signaling quasi-long range superfluid behavior also at finite $\delta$.

A careful analysis of the oscillatory character of ${\cal C}_{\ell\ell'=L/2}$  can be 
done by means of the  Fourier transform of the pairing correlator
\begin{equation}
\label{eq:ft_cooper}
        {\cal C}(k_m,k_{m'})=\sum_{\ell,\ell'=1}^{L}{\cal C}_{\ell\ell'}
        \varphi_m(\ell)\varphi_{m'}(\ell')\,.
\end{equation}
where $\varphi_{m}(\ell)=[2/(L+1)]^{-1/2}\sin{\left(k_m\ell\right)}$ [with $k_m=\pi m/(L+1)$,
$m=1\dots L$] are the eigenstates of the hopping term in Eq.~(\ref{eq:hubbard}). 
The mode with zero wave number is excluded from the allowed $k_m$ values due to the OBC. 
The lowest energy mode corresponds to $k_1$. The diagonal part of the matrix 
${\cal C}(k_m,k_{m'})$ will be simply denoted by 
${\cal C}(k_m)\equiv {\rm diag}\{{\cal C}(k_m,k_{m'})\}={\cal C}(k_m,k_m)$. In the bottom 
panel of Fig.~\ref{fig:two} we plot the difference 
$\Delta {\cal C}(k_m)= {\cal C}(k_m)-{\cal C}^{(0)}(k_m)$ between ${\cal C}(k_m)$ and its 
value in the noninteracting gas ({\it i.e.} at $U/t=0$), ${\cal C}^{(0)}(k_m)$~\cite{footnote}.

At $\delta=0$ ${\cal C}(k_m)$ possesses a very narrow peak at $k_1$ (see inset in the 
bottom panel of Fig.~\ref{fig:two}), signaling quasi-long-range superfluid order of the 
conventional BCS type. For a finite $\delta$, instead, 
${\cal C}(k_m)$ has a local minimum at $k_1$ and a {\it single} well-defined peak
 appears at a wave number $q_{\rm FFLO}= k_1+|k_{{\rm F}\uparrow}-k_{{\rm F}\downarrow}|$, 
where $k_{{\rm F}\sigma}=\pi N_\sigma/(L+1)$ are the spin-resolved Fermi wave numbers. 
The peak at $q_{\rm FFLO}$ in the Fourier transform of the pairing correlator, which 
is a direct consequence of the simple real-space nodal structure illustrated in the 
top panel of Fig.~\ref{fig:two}, is a clear-cut signal of FFLO pairing.

\begin{figure}
\begin{center}
\tabcolsep=0cm
\begin{tabular}{c}
\includegraphics[width=1.00\linewidth]{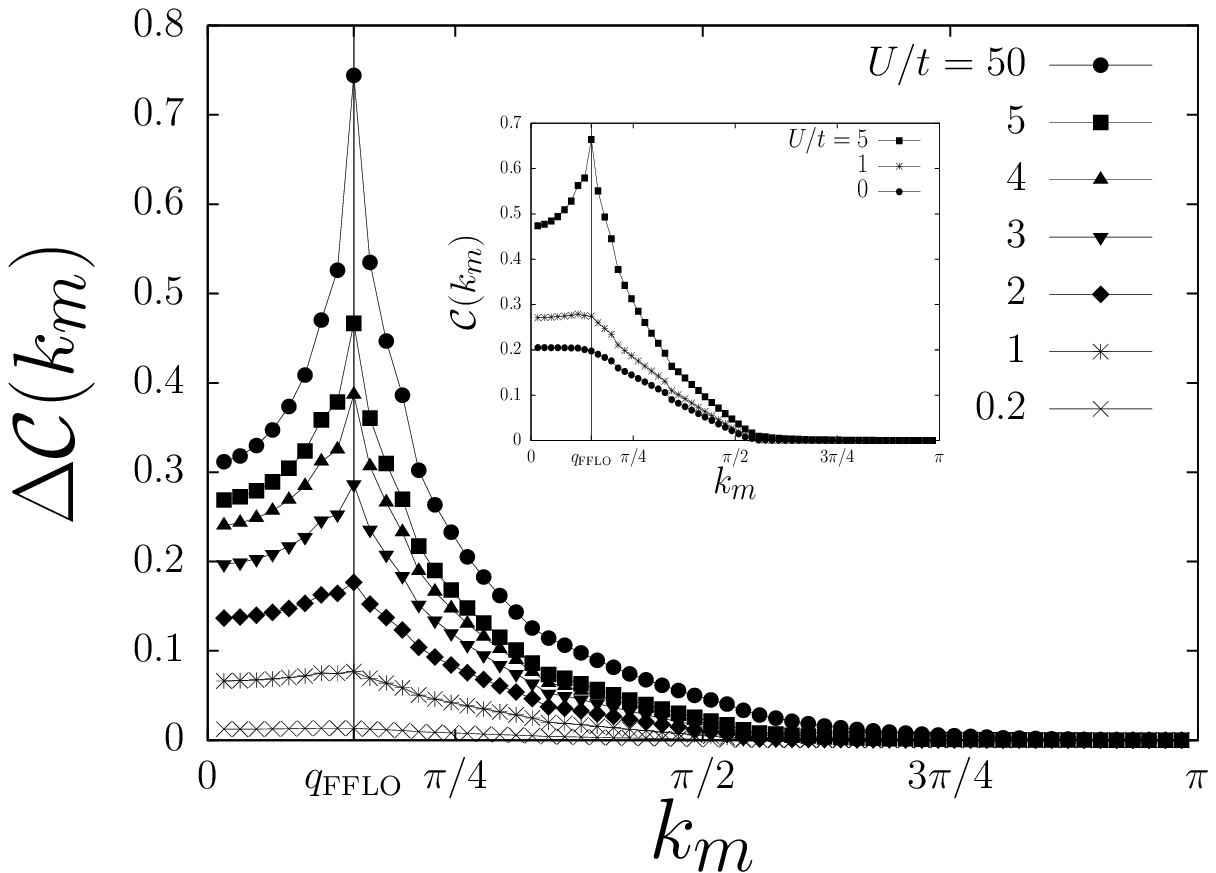}\\
\includegraphics[width=1.00\linewidth]{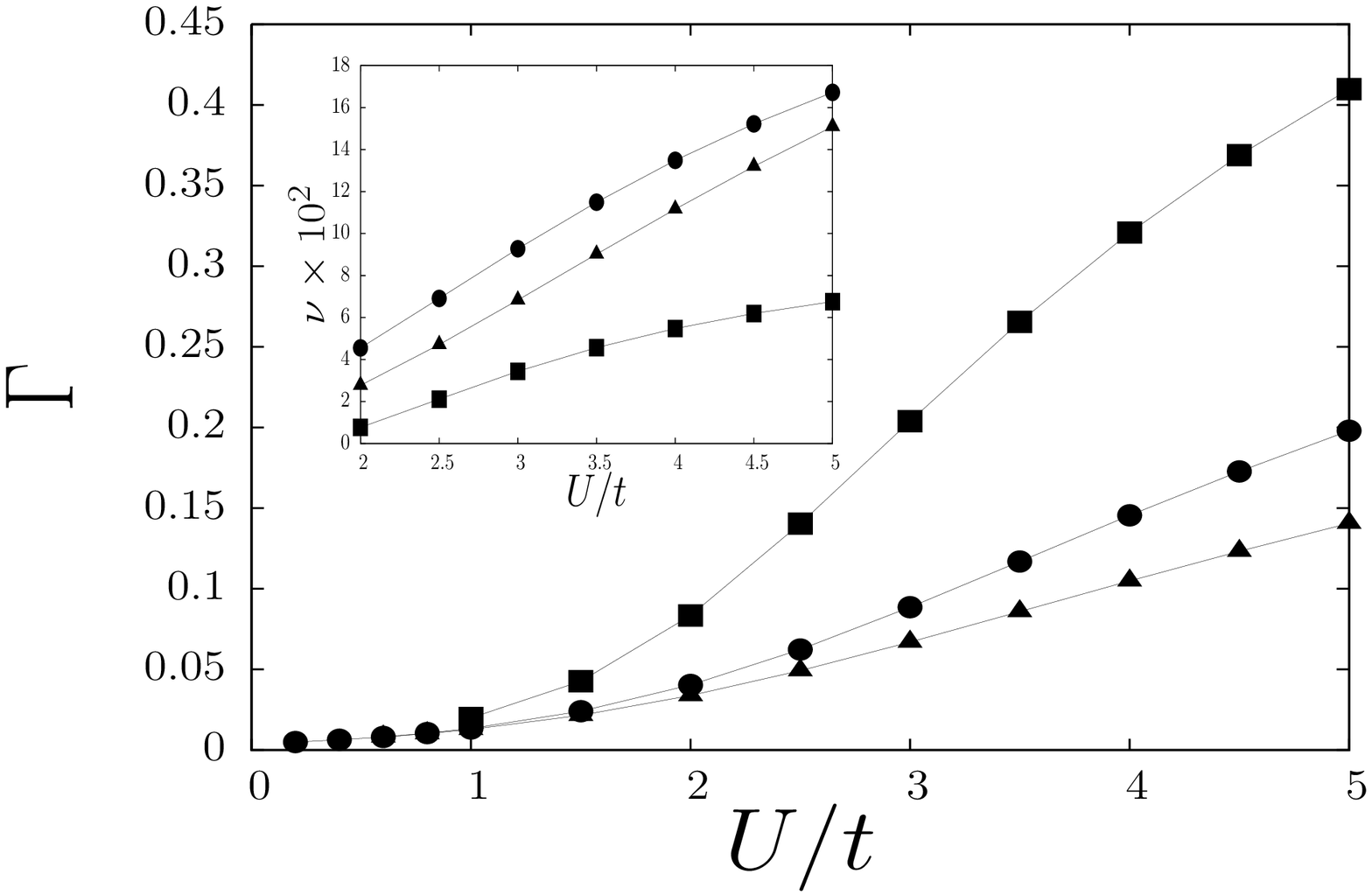}
\end{tabular}
\caption{
	Top panel: Interaction contribution to the momentum-space pairing correlator 
        $\Delta {\cal C}(k_m)$ 
	as a function of $k_m$, for $N_\uparrow=20$,  $N_\downarrow=12$ ($\delta=25\%$), and $L=60$.  
	The interaction strength $U/t$ is decreased from $50$ to $0.2$. 
	In the inset we show ${\cal C}(k_m)$ for $U/t=5, 1$, and $0$. 
	Bottom panel: The FFLO anomaly $\Gamma$ in the superfluid correlator 
        [see Eq.~(\ref{eq:anomaly})] as a function of $U/t$ for $N_\uparrow=20$ and 
	$N_\downarrow=18$ (squares, $\delta=5\%$), $N_\downarrow=12$ (circles, $\delta=25\%$), 
	and $N_\downarrow=6$ (triangles, $\delta=54\%$). 
	Inset: peak visibility $\nu$ [see Eq.~(\ref{eq:visibility})] 
	as a function of $2\leq U/t \leq 5$ for the same system parameters as in main 
        body of the figure ($\nu$ is about $30\%$ at $U/t=50$).
	The thin solid lines are just guides to the eye.
	\label{fig:three}}
\end{center}
\end{figure}

The DMRG data shown in Fig.~\ref{fig:two} refer only to a single value of $U/t=5$. We 
now turn to illustrate the dependence of the superfluid correlation functions on $U/t$.
In the top panel of Fig.~\ref{fig:three} we illustrate the dependence of $\Delta{\cal C}(k_m)$ 
on the interaction strength $U/t$ for a fixed spin polarization $\delta=25\%$. 
On decreasing $U/t$ the quasi-long-range FFLO order ({\it i.e.} the height of the peak 
at $q_{\rm FFLO}$), which is emphatically strong for large $U/t$, survives all the way 
down to the weak coupling regime. This can be quantified better by analyzing the size 
of the anomaly $\Gamma$ at $k_m=q_{\rm FFLO}$, which is measured by the difference between 
left and right (discrete) derivatives of ${\cal C}(k_m)$ evaluated at 
$q_{\rm FFLO}$,
\begin{eqnarray}\label{eq:anomaly}
\Gamma = {\cal C}(q_{\rm FFLO}+k_1)+{\cal C}(q_{\rm FFLO}-k_1)-2{\cal C}(q_{\rm FFLO})~.
\end{eqnarray}
In the bottom panel of Fig.~\ref{fig:three} we plot $\Gamma$ as a function of $U/t \leq 5$. 
In this range $\Gamma$ decreases in a smooth fashion to its noninteracting value ({\it i.e.} $\Gamma=0$) 
as $U/t$ is decreased to zero. In other words, for every finite $\delta$, ${\cal C}(k_m)$ tends 
uniformly and smoothly to its noninteracting value ${\cal C}^{(0)}(k_m)$ as $U/t$ is 
decreased towards zero. For sufficiently large values of $U/t$ the FFLO phase can also 
be characterized by  the peak visibility defined by
\begin{equation}\label{eq:visibility}
\nu= \frac{{\cal C}(q_{\rm FFLO})-{\cal C}(k_1)}{{\cal C}(q_{\rm FFLO})+{\cal C}(k_1)}\,.
\end{equation}
This quantity is plotted in an inset to the bottom panel of Fig.~\ref{fig:three}.

\begin{figure}
\begin{center}
\includegraphics[width=1.00\linewidth]{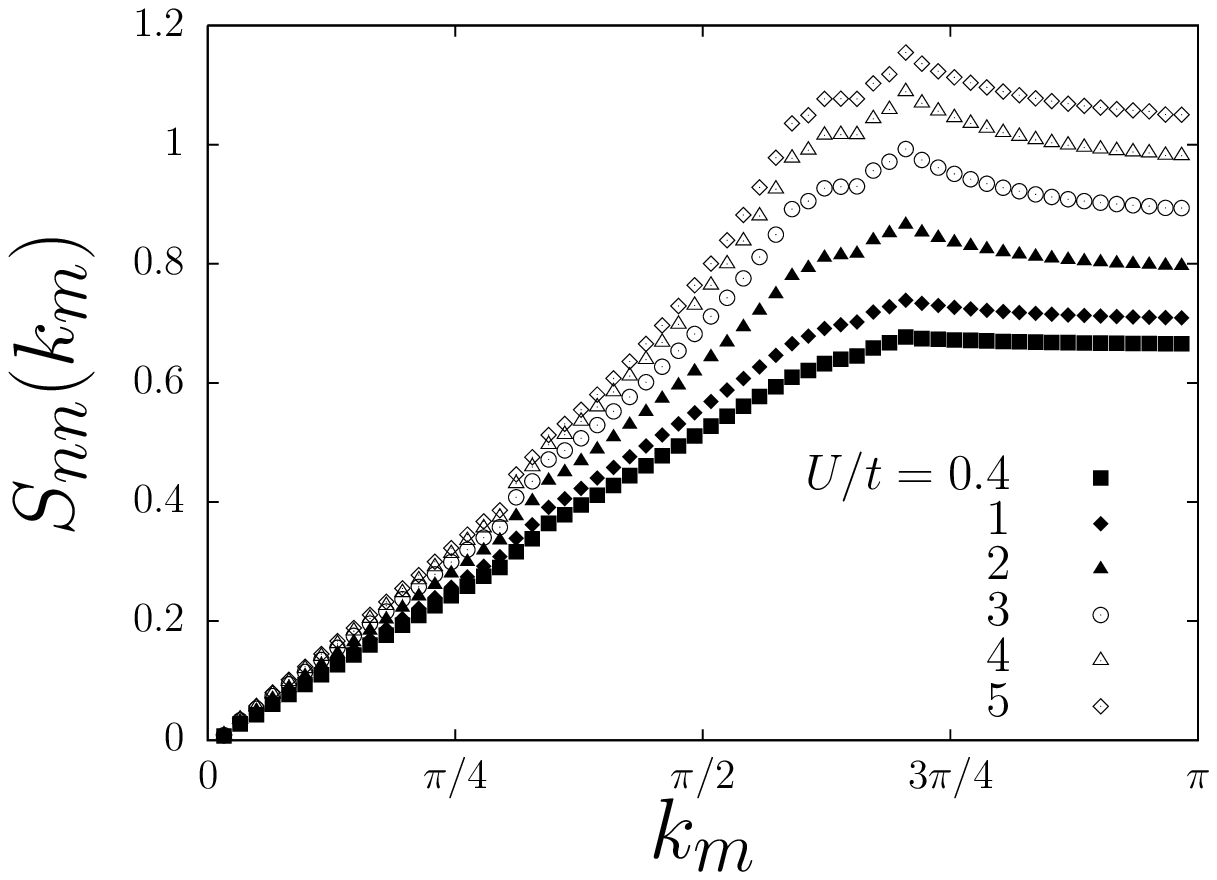}
\includegraphics[width=1.00\linewidth]{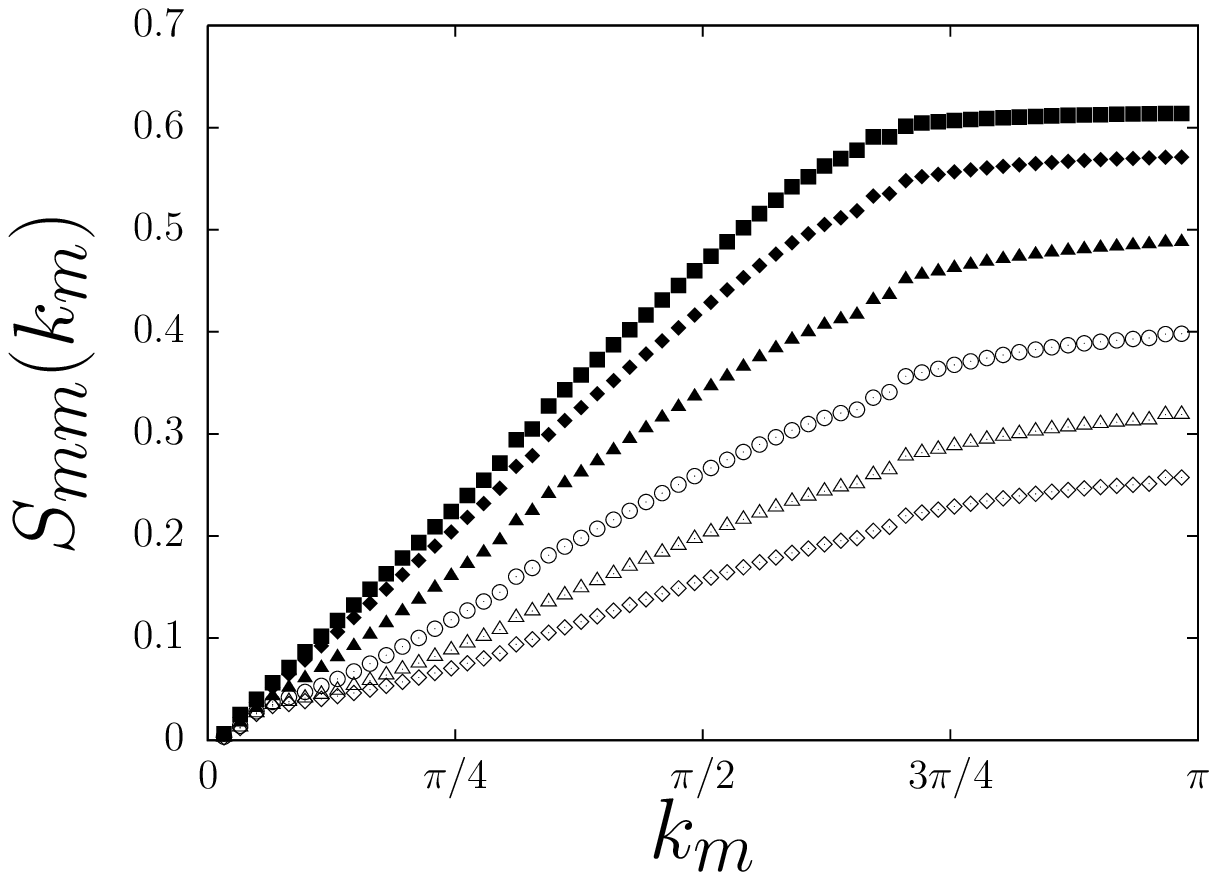}
\caption{Top panel: the density-density structure factor $S_{nn}(k_m)$ as a function of $k_m$ for 
$N_\uparrow=20$ and $N_\downarrow=18$ and different values of $U/t=0.4,1,2,3,4,$ and $5$ (from bottom to top).
Bottom panel: the spin-spin structure factor $S_{mm}(k_m)$. The symbol coding is as in the top panel.~\label{fig:four}}
\end{center}
\end{figure}

\begin{figure}
\begin{center}
\includegraphics[width=1.00\linewidth]{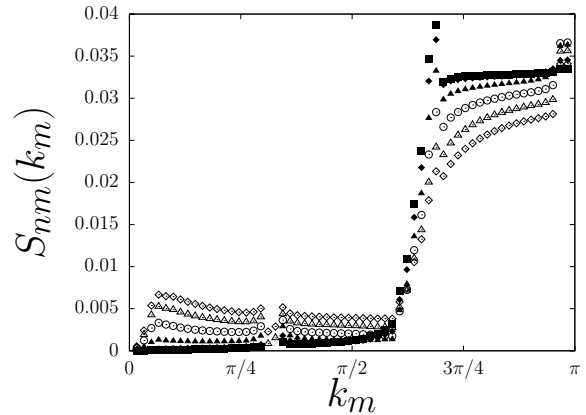}
\caption{The mixed density-spin structure factor $S_{nm}(k_m)$ as a function of $k_m$ for 
$N_\uparrow=20$ and $N_\downarrow=18$ and different values of $U/t=0.4,1,2,3,4,$ and $5$. 
The symbol coding is as in Fig.~\ref{fig:four}. Whereas the results at small $U/t$ (see, for example filled squares corresponding to $U/t=0.4$) are likely to be  somewhat affected by finite-size effects, which prevent the full development of a pairing gap, the results at larger $U$ clearly show a non-vanishing weight of $S_{nm}(k_m)$ at small $k_m$, thus indicating that spin and charge degrees of freedom are coupled at long wavelength.~\label{fig:five}}
\end{center}
\end{figure}

Before concluding, we would like to illustrate the behavior of 
the density-density, spin-spin, and mixed density-spin static structure factors, $S_{nn}(k_m)$, $S_{mm}(k_m)$, and $S_{nm}(k_m)$. 
These are defined by the sum over all frequencies of the corresponding dynamic structure factors~\cite{Giuliani_and_Vignale} that can be in principle
measured through Bragg spectroscopy or Fourier sampling of time-of-flight images~\cite{duan_prl_2006}. 
In practice, $S_{nn}(k_m)$, $S_{mm}(k_m)$, and $S_{nm}(k_m)$ are calculated from the following equations:
\begin{equation}
\left\{
\begin{array}{ll}
S_{nn}(k_m)&={\rm diag}\{{\cal F.}{\cal T.}
[\langle {\hat n}_{\ell}{\hat n}_{\ell'} \rangle-
\langle {\hat n}_{\ell} \rangle\langle {\hat n}_{\ell'}\rangle]\}\vspace{0.1 cm}\\
S_{mm}(k_m)&={\rm diag}\{{\cal F.}{\cal T.}
[\langle {\hat m}_{\ell}{\hat m}_{\ell'} \rangle-
\langle {\hat m}_{\ell} \rangle\langle {\hat m}_{\ell'}\rangle]\}\vspace{0.1 cm}\\
S_{nm}(k_m)&={\rm diag}\{{\cal F.}{\cal T.}
[\langle {\hat n}_{\ell}{\hat m}_{\ell'} \rangle-
\langle {\hat n}_{\ell} \rangle\langle {\hat m}_{\ell'}\rangle]\}
\end{array}
\right.~,
\end{equation}
where ${\hat n}_{\ell}={\hat n}_{\ell\uparrow}+{\hat n}_{\ell\downarrow}$ and 
${\hat m}_{\ell}={\hat n}_{\ell\uparrow}-{\hat n}_{\ell\downarrow}$.
In Figs.~\ref{fig:four}-\ref{fig:five} we show the dependence of
$S_{nn}(k_m)$, $S_{mm}(k_m)$, and $S_{nm}(k_m)$ on $U/t$ for a slightly asymmetric system with $N_\uparrow=20$ and $N_\downarrow=18$ ($\delta \sim 5\%$). We remind the reader that in the unpolarized $\delta=0$ case $S_{nn}(k_m)$ has a peak at 
$k_m=2k_{{\rm F}\uparrow}=2k_{{\rm F}\downarrow}$ that signals real-space atomic-density waves~\cite{gao_prl_2007,karim_pour_2007}. In the spin-polarized case, this peak splits into two peaks at $2 k_{{\rm F}\uparrow}$ and $2 k_{{\rm F}\downarrow}$. This is clearly
visible in Fig.~\ref{fig:four} in the static density-density 
structure factor (top panel), which presents a double-peak structure slightly
below $k_m = 3\pi/4$ ($2k_{{\rm F}\uparrow} \approx  2\pi/3$ and $2 k_{{\rm F}\downarrow} \approx 3\pi/5$ 
for the system parameters in this figure).
This double-peak structure is not so visible in the magnetic structure factor $S_{mm}(k_m)$, 
most likely because magnetic correlations near $2k_{{\rm F}\uparrow}$ and  $2k_{{\rm F}\downarrow}$ are still quite suppressed by
the superfluid correlations, at least in the weakly polarized case (in the unpolarized case they are completely suppressed by the
pairing gap).  From Fig.~\ref{fig:five} we note that $S_{nm}(k_m)$ is non-zero even at small $k_m$, thus indicating that spin and charge degrees of freedom are coupled at long wavelength even for a small imbalance.

\subsection*{Experimental signatures of the FFLO phase}

The most direct way to detect FFLO pairing would be 
to measure the pairing correlation function ${\cal C}_{\ell \ell'}$. We would like to remark here that
this correlation function is, at least in principle, measurable via interferometric schemes~\cite{carusotto} in which 
two atomic wave packets are coherently extracted from the gas at different positions and then are mixed by a matter-wave beam splitter. The atom counting statistics in the beam splitter output channels has been shown~\cite{carusotto} to reflect the spatial dependence of  ${\cal C}_{\ell \ell'}$. 

The oscillations of the pairing correlations will also leave a detectable signature in the noise correlations~\cite{noise,greiner_shot_noise},
$G_{\uparrow \downarrow}({\bm k},{\bm k}') =
\langle {\hat n}_{{\bm k}, \uparrow} {\hat n}_{{\bm k}', \downarrow}\rangle -\langle {\hat n}_{{\bm k}, \uparrow}\rangle \langle {\hat n}_{{\bm k}', \downarrow}\rangle $, where ${\hat n}_{{\bm k}, \sigma}$ measures
the number of fermions with momentum ${\bm k}$ and spin $\sigma$ in a time-of-flight experiment.
With increasing spin-polarization, in fact, the peak at  ${\bm k} = -{\bm k}'  = (k,0,0)$~\cite{mathey,noack} 
[here $(1,0,0)$ is the direction along the
axis of the 1D system] will shift to a finite relative momentum (see {\it e.g.} the work by Yang in Ref.~\onlinecite{population_imbalance_theory} and the very recent DMRG calculation by L\"uscher {\it et al.}~\cite{noack_last}). 

However, it is worth pointing out that the strength of noise signal in a strictly-1D system will 
be strongly affected by finite-size and temperature effects. This is because in 1D the order is not long-range but quasi-long range,
and therefore the slowest decay exhibited by correlations (like the pairing correlations) is a  power law. 
Thus, in order to enhance the strength of the experimental signal for FFLO, it would be desirable to couple many 1D systems, 
as in a tight 2D optical lattice (arrays of ``atomic quantum wires'')~\cite{references_on_2DOL}, so that the quasi-long
range  FFLO order can become true long-range order. The phase diagram of many coupled 1D systems has been worked out in Ref.~\onlinecite{yang}, where the author found that, at small polarization $\delta$, true long-range 3D FFLO order will occur when the Luttinger liquid parameter for the charge excitations, $K_{\rho}$, is larger than $3/2$. In such a case the low temperature properties of the system are dominated by hopping of pairs ({\it i.e.} Josephson coupling) rather than by single-particle hopping (the latter would turn the system into
an anisotropic Fermi liquid, which could in turn become unstable to the FFLO state under appropriate conditions~\cite{huse_etal}). 

However,  the analysis of Ref.~\onlinecite{yang} assumed the transition from the unpolarized to the polarized case to belong the 
the commensurate-incommensurate universality class. This assumption implicitly neglects the coupling between charge and spin
degrees of freedom at low energies, which is known to modify the behavior of physical observables at the transition~\cite{Woynarovich, Frahm07}.  
In this work this coupling has been demonstrated to exist also at long wavelengths by an explicit numerical evaluation 
of the mixed static structure factor $S_{nm}(k_m)$ in a weakly polarized system (see Fig.~\ref{fig:five}). Thus, the phase diagram depicted in Fig.~1 of Ref.~\onlinecite{yang} seems not appropriate for coupled 1D Hubbard (or Gaudin-Yang~\cite{Guan07}) models, and the transition to long-range order will not take place in general for $K_{\rho} = 3/2$ and may in general depend on the system parameters 
({\it i.e.} $U/t$ and the lattice filling for the Hubbard model).

To the best of our knowledge, a quantitative phase diagram of coupled 1D systems lacking spin-charge separation 
has not yet been calculated. Furthermore, it is worth noticing that in cold
atomic systems with short range interactions, another important factor must be taken into account, namely the relative strength of the pair hopping when compared to the single-particle hopping. The strength of the latter is given by 
$t_{\perp}$ ($t_{\perp} \ll \varepsilon_{\rm F}$, where $\varepsilon_{\rm F}$ is the Fermi energy, 
for the analysis based on coupled Tomonaga-Luttinger liquids to hold), where
$t_{\perp}$ is the hopping amplitude between two neighboring 1D systems. However, in absence of long-range
interactions pair hopping can be only generated by (virtual) single-particle tunneling events, which at lowest
order, yield a (Josephson) coupling strength of order $t^2_\perp/\Delta_{\sigma}$, where $\Delta_\sigma$ is the spin gap. 
The phase diagram  predicted in  Ref.~\onlinecite{yang} is the result of a calculation which only compares the scaling
dimensions of the pair hopping and single-particle hopping operators, and thus does not take into account the
microscopic details of the coupling between 1D systems. Thus, the stabilization of  
FFLO long-range order by a weak coupling between 1D systems in the FFLO phase (the extreme anisotropic limit that
could not be accessed by the authors of Ref.~\onlinecite{huse_etal}) does not seem easily achievable. In turn,
the most likely scenario for arbitrary polarization is that the single-particle hopping will control the physics 
at low temperatures,  and the system will behave as a spin-polarized normal Fermi liquid, which, in turn, 
could become unstable towards 3D FFLO ordering~\cite{huse_etal} under appropriate conditions.

\section{Conclusions}
\label{sect:summary}

In summary, we have shown how ultracold spin-polarized two-component Fermi gases  confined in 1D optical lattices are FFLO superfluids whose pairing correlation functions  are characterized by a power-law decay and a simple nodal structure. However, we have also shown that charge and spin degrees of freedom appear to be coupled already for a small value of the spin polarization. Finally, we have commented on the impact of this coupling on the detectability of true long-range FFLO order arising from Josephson coupling between 1D systems.

\acknowledgments

We acknowledge useful discussions with P. Calabrese, I. Carusotto, M. K\"ohl, and A.H. MacDonald. 
M.P. gratefully acknowledges the hospitality of the Donostia International Physics Center and of the 
Department of Physics of the Zhejiang Normal University during the final stages of this work. 
M.A.C. thanks A. Nersesyan for drawing attention to Refs.~\onlinecite{Woynarovich}. 
This work was partly supported by PRIN-MIUR and by the Academy of Finland (project number 115020).
The DMRG calculations have been made using the DMRG code released
within the ``Powder with Power'' Project (www.qti.sns.it).


\begin{thebibliography}{77}
\bibitem{footnote}
	${\cal C}^{(0)}(k_m)$ is calculated from the noninteracting real-space correlator 
	${\cal C}^{(0)}_{\ell\ell'}=\langle {\hat c}^\dagger_{\ell\uparrow}{\hat c}_{\ell'\uparrow}\rangle
	\langle {\hat c}^\dagger_{\ell\downarrow}{\hat c}_{\ell'\downarrow}\rangle$, 
	with $\langle {\hat c}^\dagger_{\ell\sigma}{\hat c}_{\ell'\sigma}
        \rangle=\sum_{n=1}^{N_\sigma}\varphi_{k_n}(\ell)\varphi_{k_n}(\ell')$.
\end{thebibliography}
\end{document}